\documentclass[prl,amssymb,twocolumn]{revtex4}
\usepackage{graphicx}
\begin{document}

{\large \bf Resonant-pulse operations on the buried donor
charge qubits in semiconductors}

\vskip 6mm

\centerline{L. A. Openov}

\vskip 4mm

{\it Moscow Engineering Physics Institute (State
University), 115409 Moscow, Russia}

\vskip 6mm

\begin{quotation}

A new scheme is proposed for rotations of a double-donor charge qubit whose
logical states are defined by the two lowest energy states of a single
electron localized around one or another donor. It is shown that making use
of the microwave pulses tuned to the resonance with an auxiliary excited
molecular level allows for implementation of various one-qubit operations in
very short times. Decoherence effects are analyzed by the example of the
P$_2^+$:Si system and shown to be weak enough for experimental realization of
this scheme being possible.

\end{quotation}

\vskip 6mm


\vskip 8mm

Semiconductor-based devices seem to be very promising for a scalable
quantum computing technology \cite{QC}. The qubits can be encoded, e. g.,
in the nuclear or electron spin states \cite{Kane,Vrijen}. Although the
relatively long coherence times make the spin-based qubits good candidates
for quantum computation, the single-spin measurement still remains a
significant challenge \cite{Vrijen,Kane2}. The charged-based qubits in
semiconductors are currently discussed as well, their logical states being
encoded in the orbital degrees of freedom of an electron occupying the
quantum-dot structure \cite{Barenco,Tanamoto,Valiev,Hayashi}.
Coherent oscillations
of the double-dot qubit have been observed \cite{Hayashi}, and the readout
schemes have been proposed \cite{DiCarlo}. In spite of the fact that
decoherence of the charge-based qubits \cite{Petta}
is much stronger than of their
spin-based counterparts, the charge-based qubits are nevertheless believed
to be realizable at the present technological level due to their short
operation times.

In view of the technical difficulties concerning manufacturing of the
quantum dots with predetermined characteristics, it may appear more
reasonable to make use of natural atoms (instead of "artificial" ones) as
the localization centers for the electrons carrying qubits. Recent advances
in atomically precise placement of single dopants in semiconductors
\cite{Schofield,Dzurak} make possible the construction of solid-state atomic
qubits. Hollenberg {\it et al.} proposed a two-atom scheme where the charge
qubit consists of two dopant atoms beneath the semiconductor surface
\cite{Dzurak,Hollenberg}. One of
the donors is singly ionized, and the logical states are formed by the lowest
two energy states of the remaining valence electron localized at the left or
the right donor, $|0\rangle = |L\rangle$ and $|1\rangle = |R\rangle$,
see Fig. 1. Initialization and readout of the qubit are facilitated by a
single electron transistor. The surface electrodes are used to control the
qubit through the adiabatic variations of the donor potentials.

\begin{figure}[h]
\includegraphics[width=0.4\textwidth,height=5cm]{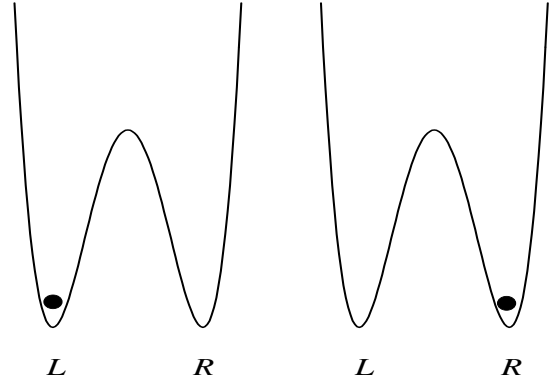}
\caption{The logical states $|0\rangle=|L\rangle$ and $|1\rangle=|R\rangle$
of the buried donor charge qubit.}
\label{Fig1}
\end{figure}

Here we propose an alternative scheme for rotations of the buried donor
charge qubit. Instead of applying biases to the surface gates, the qubit is
driven by two microwave pulses tuned to the resonances between the localized
states $|L\rangle$, $|R\rangle$ and one of the excited molecular states
delocalized over
the double-donor structure. We present the analytical solution for the
three-level model of the unitary electron evolution and show that, depending
on the specific values of frequencies, phases, amplitudes, and durations of
the pulses, various one-qubit operations can be implemented in times as short
as $\sim 1$ ps, i. e., more than one order of magnitude shorter than in the
original proposal \cite{Dzurak,Hollenberg}. At such times, decoherence is
sufficiently weak, making it possible at least to investigate the small-scale
devices and thus to demonstrate the experimental feasibility of the scheme.

We consider a singly ionized pair of dopant atoms in a semiconductor host by
the example of the P$_2^+$:Si system. The Hamiltonian of the remaining
valence electron is
\begin{equation}
\hat{H}_0=\sum_{n=1} E_n |\chi_n\rangle \langle \chi_n|~,
\label{H0}
\end{equation}
where $E_n$ and $|\chi_n\rangle$ are, respectively, the one-electron
eigenenergies and eigenstates of the diatomic ion. In order to avoid the
sophisticated numerical calculations \cite{Kettle}, we ignore the
conduction-band anisotropy, the inter-valley terms, the surface effects,
etc., and make use of a simple effective mass approximation that is commonly
used for the semi-quantitative considerations \cite{Barrett,Fedichkin}. Then,
in the case that the two donors are equivalent, the problem reduces to that
for the hydrogen-like molecular ion with the effective Bohr radius
$a_B^*\approx$ 3 nm and the effective Hartree unit of energy
$E^*=e^2/\varepsilon a_B^*\approx$ 40 meV, where $\varepsilon\approx 12$
is the dielectric constant for silicon. The lowest states $|\chi_1\rangle$
and $|\chi_2\rangle$ are the molecular states $1s\sigma_g$ and
$2p\sigma_u$, whose wave functions are, respectively, symmetric and
antisymmetric about the midpoint of the line joining the two donors.
At large donor separation, $R>>a_B^*$, one has
$E_1\approx E_2\approx -E^*/2$, while the difference $\Delta E_{21}=E_2-E_1$
is exponentially small \cite{Herring},
$\Delta E_{21}/E^*=4xe^{-x-1}\left[1+O(1/x)\right]$, where $x=R/a_B^*$.
The excited state $|\chi_3\rangle$ with the energy $E_3\approx -E^*/8$ is
well separated from the states $|\chi_1\rangle$ and $|\chi_2\rangle$ by
$\Delta E_{31}\approx\Delta E_{32}\approx 3E^*/8\approx 15$ meV.
At $R/a_B^*>6$ this is the molecular state $3d\sigma_g$. If the
qubit is biased by the static gate voltages, so that the difference $E_R-E_L$
in the energies of the lowest states $|L\rangle$ and $|R\rangle$ localized
at different donors greatly
exceeds the value of $\Delta E_{21}$, then the states
$|L\rangle$ and $|R\rangle$ are good approximations to the lowest
eigenstates, and
\begin{equation}
\hat{H}_0=E_L|L\rangle\langle L|+E_R|R\rangle\langle R|+
\sum_{n=3} E_n |\chi_n\rangle\langle \chi_n|~.
\label{H01}
\end{equation}

Let the qubit interact with an external electromagnetic field
${\bf E}(t)={\bf E_{01}}(t)\cos(\omega_L t)+
{\bf E_{02}}(t)\cos(\omega_R t+\phi)$ that has two components oscillating at
frequencies $\omega_L=(E_{3}-E_L)/\hbar$ and $\omega_R=(E_{3}-E_R)/\hbar$,
where ${\bf E_{01}}(t)$ and ${\bf E_{02}}(t)$ are the slowly varying
envelopes. Then the Hamiltonian becomes
\begin{equation}
\hat{H}(t)=\hat{H}_0+\hat{V}(t)~.
\label{H}
\end{equation}
The interaction term $\hat{V}(t)$ is
\begin{equation}
\hat{V}(t)={\bf E}(t)\biggl[ {\bf d}_L|\chi_3\rangle \langle L|+
{\bf d}_R|\chi_3\rangle \langle R| + h.c. \biggr] ~,
\label{V}
\end{equation}
where ${\bf d}_{L,R}=\langle \chi_3|-e{\bf r}|L,R\rangle$ are the electric
dipole moments for the transitions
$|L,R\rangle \rightleftharpoons|\chi_3\rangle$.
In the resonant approximation \cite{NOTE} one has
\begin{eqnarray}
&& \hat{V}(t)=\frac{1}{2}e^{-i\omega_L t}\lambda_L(t)
|\chi_3\rangle \langle L|
\nonumber
\\ && + \frac{1}{2}e^{-i\omega_R
t-i\phi}\lambda_R(t) |\chi_3\rangle \langle R| + h. c.~,
\label{V(t)}
\end{eqnarray}
where $\lambda_L(t)={\bf E_{01}}(t){\bf d}_L$ and
$\lambda_R(t)={\bf E_{02}}(t){\bf d}_R$.
Here we restrict ourselves to the rectangular pulse shapes, so
that both ${\bf E_{01}}(t)$ and ${\bf E_{02}}(t)$ are constant at
$0<t<\tau_{op}$ and zero elsewhere.

It is straightforward to solve the non-stationary Schr\"{o}dinger equation
\begin{equation}
i\hbar\frac{\partial |\Psi(t)\rangle}{\partial t}=\hat{H}(t)|\Psi(t)\rangle
\label{Sch(t)}
\end{equation}
for the state vector
\begin{eqnarray}
&& |\Psi(t)\rangle = C_L(t)e^{-iE_L t/\hbar}|L\rangle +
C_R(t)e^{-iE_R t/\hbar}|R\rangle
\nonumber
\\ &&
+C_3(t)e^{-iE_3 t/\hbar}|\chi_3\rangle
\label{Psi(t)}
\end{eqnarray}
and to find the coefficients $C_L(t)$, $C_R(t)$, and $C_3(t)$ provided that
$|\Psi(0)\rangle = \alpha|L\rangle + \beta|R\rangle$:
\begin{eqnarray}
&&C_L(t)=\alpha\left [
1-\frac{2|\lambda_L|^2}{|\lambda_L|^2+|\lambda_R|^2} \sin^2(\Omega
t)\right ]
\nonumber
\\ && -\beta\frac{2\lambda_L^*\lambda_R e^{-i\phi}}
{|\lambda_L|^2+|\lambda_R|^2}\sin^2(\Omega t)~,
\nonumber
\\ && C_R(t)=-\alpha\frac{2\lambda_L\lambda_R^* e^{i\phi}}
{|\lambda_L|^2+|\lambda_R|^2}
\sin^2(\Omega t)
\nonumber
\\ && +\beta\left [ 1-\frac{2|\lambda_R|^2}
{|\lambda_L|^2+|\lambda_R|^2}\sin^2(\Omega t) \right ] ~,
\nonumber
\\ && C_3(t)=-i\frac{\alpha\lambda_L+\beta\lambda_R e^{-i\phi}}
{\sqrt{|\lambda_L|^2+|\lambda_R|^2}}\sin(2\Omega t)~,
\label{C(t)}
\end{eqnarray}
where $\Omega=\sqrt{|\lambda_L|^2+|\lambda_R|^2}/4\hbar$. From
Eq. (\ref{C(t)}) one can see that at $t=\tau_{op}=\pi k/2\Omega$
($k$ is a positive integer) the coefficient $C_3$ vanishes, and the state
vector $|\Psi(t)\rangle$ remains in the qubit subspace
$\{|L\rangle,|R\rangle\}$. So, the auxiliary excited state $|\chi_3\rangle$
assists the qubit evolution by means of the electron transfer between the
states $|L\rangle$ and $|R\rangle$ as the driving field is on but remains
unpopulated after the field is off \cite{Openov}.

It follows from Eqs. (\ref{Psi(t)}) and (\ref{C(t)}) that the relative phase
shift operation,
\begin{equation}
|\Psi(\tau_{op})\rangle = e^{-iE_L\tau_{op}/\hbar}
\biggl[\alpha|L\rangle +
\beta e^{-i(E_R-E_L)\tau_{op}/\hbar}|R\rangle \biggr] ~,
\label{Psi(tau_op)1}
\end{equation}
is implemented at $\tau_{op}=\pi k/\Omega$. The quantum NOT operation,
\begin{equation}
|\Psi(\tau_{op})\rangle = \pm e^{-iE_L\tau_{op}/\hbar-i\phi}
\biggl[\beta|L\rangle + \alpha|R\rangle \biggr]~,
\label{Psi(tau_op)2}
\end{equation}
is realized at $\tau_{op}=\pi (2k-1)/2\Omega$ if $\lambda_L=\mp\lambda_R$
and $\phi=\pi n + (E_R-E_L)\tau_{op}/2\hbar$ ($n$ is an integer). The
Hadamard transformation,
\begin{equation}
|\Psi(\tau_{op})\rangle = \pm e^{-iE_L\tau_{op}/\hbar}
\left[\frac{\alpha+\beta}{\sqrt{2}}|L\rangle +
\frac{\alpha-\beta}{\sqrt{2}}|R\rangle\right]~,
\label{Psi(tau_op)3}
\end{equation}
is performed at $\tau_{op}=\pi (2k-1)/2\Omega$ if
$(E_R-E_L)\tau_{op}/\hbar=2\pi m$ ($m$ is a positive integer). The plus
sign in Eq. (\ref{Psi(tau_op)3}) corresponds to $\phi=2\pi n$ and
$\lambda_L=-\lambda_R(\sqrt{2}-1)$ or
$\phi=\pi(2n+1)$ and $\lambda_L=\lambda_R(\sqrt{2}-1)$, and the minus sign
corresponds to $\phi=2\pi n$ and $\lambda_L=\lambda_R(\sqrt{2}+1)$ or
$\phi=\pi(2n+1)$ and $\lambda_L=-\lambda_R(\sqrt{2}+1)$.

For the field amplitudes $E_0\sim 1$ V/cm, the operation time is
$\tau_{op}\sim 1/\Omega\sim \hbar/|\lambda_{L,R}|\sim \hbar/ea_B^*E_0\sim 1$
ns. Increase in the pulse intensity will cause the value of $\tau_{op}$ to
decrease down to the picosecond time scale, so that the value of $\tau_{op}$
can be made much shorter than in the case that the qubit is manipulated by
adiabatically varying the potentials of the surface gates \cite{Hollenberg}.
Note that the energies $E_L$ and $E_R$ should be sufficiently different from
each other, $E_R-E_L\sim 1$ meV, in order the qubit rotations could be
implemented in times $\tau_{op}\sim 1$ ps.

The uncontrolled interaction of the quantum system with its environment leads
to entanglement between the states of the system and the environmental
degrees of freedom. This disturbs the unitary evolution of the system and
results in the loss of coherence. There are various sources of decoherence
in solids. For the buried donor charge qubit decoherence due to the phonon
emission/absorption processes was studied in Refs. \cite{Hollenberg,Barrett}
and found to be much weaker than decoherence due to both Nyquist-Johnson
voltage fluctuations in the surface electrodes and 1/f noise from the
background charge fluctuations. Contrary to this statement, here we show that
phonons are the main cause for decoherence at short operation times. For
simplicity, we consider the qubit at zero temperature and assume isotropic
acoustic phonons with the linear dispersion law, $\omega_{{\bf q}}=sq$,
where $s$ is the speed of sound.

Electron-phonon coupling in confined systems is described by the Hamiltonian
\cite{Brandes}
\begin{equation}
\hat{H}_{ep}=\sum_{\bf q}\lambda({\bf q})\hat{\rho}({\bf q})
\left[\hat{b}_{\bf q}^+ +\hat{b}_{-{\bf q}}^{}\right]~,
\label{Hep}
\end{equation}
where
$\hat{b}_{\bf q}^+$ and $\hat{b}_{\bf q}^{}$ are, respectively, the operators
of creation and annihilation of a phonon with the wave vector ${\bf q}$,
$\hat{\rho}({\bf q})=\int{d{\bf r} e^{i{\bf qr}}\hat{\rho}({\bf r})}$
is the Fourier transform of the electron density operator
$\hat{\rho}({\bf r})=\sum_{mn}\Psi_m^*({\bf r})\Psi_n^{}({\bf r})
|m\rangle\langle n|$, and $\lambda({\bf q})$ is the microscopic
electron-phonon interaction matrix element that can be expressed in terms of
the deformation potential $D$ and the density of the crystal $\rho$ as
$\lambda({\bf q})=qD\left(\hbar/2\rho\omega_{\bf q}\Omega\right)^{1/2}$,
with $\Omega$ being the normalizing volume.

Since at $R>>a_B^*$ the overlap $\langle L|R\rangle$ between the orbitals
$\langle{\bf r}|L,R\rangle=(\pi(a_B^*)^3)^{-1/2}
\exp(-|{\bf r}-{\bf r}_{L,R}|/a_B^*)$, where
${\bf r}_{L,R}=\mp(R/2){\bf e}_x$
are the donor coordinates, is negligibly small,
the transitions $|L\rangle\rightleftharpoons |R\rangle$ are suppressed, and
decoherence of the lowest states $|L,R\rangle$ is entirely due to dephasing
processes, so that the diagonal elements of the density matrix remain
unchanged, while the off-diagonal elements are \cite{Fedichkin}
\begin{equation}
\rho_{LR}(t)=\rho_{LR}(0)e^{-B^2(t)+i(E_R-E_L)t/\hbar}~,
\label{rhoLR}
\end{equation}
where
\begin{equation}
B^2(t)=\frac{8}{\hbar^2}\sum_{{\bf q}}
\frac{|g({\bf q})|^2}{\omega_{{\bf q}}^2}
\sin^2\left(\frac{\omega_{{\bf q}}t}{2}\right)~,
\label{B2(t)}
\end{equation}
and
\begin{eqnarray}
&& g({\bf q})=\frac{\lambda({\bf q})}{2}
\left[\langle L|e^{i{\bf qr}}|L\rangle - \langle R|e^{i{\bf qr}}|R\rangle
\right]
\nonumber
\\ && =-i\lambda({\bf q})\frac{\sin(q_xR/2)}{\bigl[1+(qa_B^*)^2/4\bigr]^2}~.
\label{g(q)}
\end{eqnarray}
At $\tau_{op}>a_B^*/s$ one has
\begin{equation}
B^2(\tau_{op})=\frac{D^2}{3\pi^2\rho\hbar s^3(a_B^*)^2}~,
\label{B2(top)}
\end{equation}
so that the spectral function (\ref{B2(t)}) appears to be a material
constant, being about $6\cdot10^{-3}$ for P$_2^+$:Si, and the error rate
(i. e., the error generated during the operation time) is \cite{Fedichkin}
\begin{equation}
D(\tau_{op})=\frac{1}{2}\left[1-e^{-B^2(\tau_{op})}\right]
\approx 3\cdot 10^{-3}~.
\label{D1}
\end{equation}

Since the excited level $|\chi_3\rangle$ becomes temporarily populated during
the resonant-pulse operations on the P$_2^+$:Si qubit, the phonon emission
processes $|\chi_3\rangle\rightarrow|L,R\rangle$ also contribute to
decoherence at $T=0$. For the double donor orientation along the
$x$-axis one has $|\chi_3\rangle\approx
\bigl[|2S\rangle_L-|2P_x\rangle_L+|2S\rangle_R+|2P_x\rangle_R\bigr]/2$
at $R>>a_B^*$ and $E_R-E_L<<E_3-E_{L,R}$. Neglecting the exponentially small
overlap between the localized atomic-like orbitals
$\langle{\bf r}|L,R\rangle$,
$\langle{\bf r}|2S\rangle_{L,R}=(8\pi (a_B^*)^3)^{-1/2}
(1-|{\bf r}-{\bf r}_{L,R}|/2a_B^*)\exp(-|{\bf r}-{\bf r}_{L,R}|/2a_B^*)$, and
$\langle{\bf r}|2P_x\rangle_{L,R}=(32\pi (a_B^*)^5)^{-1/2}
(x-x_{L,R})\exp(-|{\bf r}-{\bf r}_{L,R}|/2a_B^*)$ centered at different
donors, we have
\begin{equation}
\langle \chi_3|e^{i{\bf qr}}|L,R\rangle=
2\sqrt{2}\frac{(qa_B^*)^2\mp i\frac{3}{2}(q_xa_B^*)}
{\left[\frac{9}{4}+(qa_B^*)^2\right]^3}e^{\mp iq_xR/2}~,
\label{TR_LR}
\end{equation}
so that the relaxation rate at $T=0$ is \cite{Fermi}
\begin{eqnarray}
&& \Gamma=\frac{2\pi}{\hbar}\sum_
{{\bf q},L,R}|\lambda({\bf q})|^2|\langle \chi_3|e^{i{\bf qr}}|L,R\rangle|^2
\delta(\hbar\omega_0-\hbar\omega_{\bf q})
\nonumber
\\ && \approx \frac{8D^2}{\pi\rho\hbar s^2(a_B^*)^3}
(q_0a_B^*)^5 \frac{ \frac{3}{4}+(q_0a_B^*)^2}
{\left[\frac{9}{4}+(q_0a_B^*)^2\right]^6} ~,
\label{Gamma}
\end{eqnarray}
where $\hbar\omega_0=\hbar q_0 s=E_3-E_L\approx E_3-E_R\approx 3E^*/8$. From
Eq. (\ref{Gamma}) one has $\Gamma\approx 3\cdot 10^7$ s$^{-1}$ for
P$_2^+$:Si. We see that at
$\tau_{op}<100$ ps the error rate due to the phonon emission
processes \cite{Fedichkin},
\begin{equation}
D(\tau_{op})=1-e^{-\Gamma\tau_{op}}~,
\label{D2}
\end{equation}
is lower than
the value of $D(\tau_{op})$ due to dephasing. So, the phonon-induced error
rate at $T=0$ and short operation times is
$D(\tau_{op})\approx 3\cdot 10^{-3}$. At finite temperatures, such that
$k_B T > \hbar\omega_0$, where $\hbar\omega_0=\hbar s/a_B^*\approx 2$ meV
for dephasing processes and $\hbar\omega_0=E_4-E_3\approx E_R-E_L\sim 1$ meV
for the processes of the phonon absorption by an electron temporarily
occupying the excited state, the error rate increases by a factor of
$\sim k_B T / \hbar\omega_0$, i. e., changes slightly at $T<10$ K.

The error rate due to phonons should be compared to the error rates due to other sources of
decoherence. The lowest bounds for the decoherence times associated with the
Johnson noise from the gates and the environmental charge fluctuations are
\cite{Hollenberg,Dzurak,Barrett} $\tau\sim 1$ $\mu$s and
$\tau\sim 1$ ns, respectively, so that the corresponding error rates
$D(\tau_{op})=1-\exp(-\tau_{op}/\tau)$ do not exceed that due to phonons at
$\tau_{op}<(1\div 10)$ ps. Hence, the performance of the buried donor charge
qubit appears to be limited primarily by the electron-phonon interaction.

In conclusion, we proposed a scheme for fast rotations of the buried donor
charge qubit through an auxiliary-state-assisted electron evolution under
the influence of the resonant microwave pulses. This scheme allows for
implementation of the one-qubit operations in times as short as
$\tau_{op}\sim 1$ ps. By the example of the P$_2^+$:Si qubit, we have shown
that dephasing due to acoustic phonons is the main source of decoherence
at $T<10$ K and operation times $\tau_{op}=(1\div 10)$ ps. The error rate is
about $3\cdot 10^{-3}$, i. e., greater than the fault-tolerance threshold
for quantum computation \cite{DiVincenzo}
but low enough for coherent qubit manipulation being
possible, at least in the proof-of-principle experiments on one-qubit
devices. The coupling of the double-donor qubits via the Coulomb interaction
allows, in principle, to realize the conditional gates. It would be also
worthwhile to search for other materials and/or doping elements for the
buried donor charge qubits, in order to weaken the decoherence effects.
Although we restricted ourselves to rectangular shapes of the resonant
pulses, our consideration can be generalized to other pulse shapes

Discussions with A. V. Tsukanov, L. Fedichkin, and M. S. Litsarev are
gratefully acknowledged.


\vskip 2mm

\begin{thebibliography}{5}

\bibitem{QC}

M. A. Nielsen and I. L. Chuang, {\it Quantum Computation and
Quantum Information} (Cambridge University Press, Cambridge, 2000);
A. Olaya-Castro and N. F. Johnson, e-print quant-ph/0406133,
to appear in "Handbook of Theoretical and Computational Nanotechnology".

\bibitem{Kane} B. E. Kane, Nature {\bf 393}, 133 (1998).

\bibitem{Vrijen} R. Vrijen, E. Yablonovitch, K. Wang, H. W. Jiang,
A. Balandin, V. Roychowdhury, T. Mor, and D. DiVincenzo,
Phys. Rev. A {\bf 62}, 012306 (2000).

\bibitem{Kane2} B. E. Kane, N. S. McAlpine, A. S. Dzurak, R. G. Clark,
G. J. Milburn, He Bi Sun, and H. Wiseman, Phys. Rev. B {\bf 61}, 2961 (2000).

\bibitem{Barenco} A. Barenco, D. Deutsch, A. Ekert, and R. Jozsa,
Phys. Rev. Lett. {\bf 74}, 4083 (1995).

\bibitem{Tanamoto} T. Tanamoto, Phys. Rev. A {\bf 61}, 022305 (2000).

\bibitem{Valiev} L. Fedichkin, M. Yanchenko, and K. A. Valiev,
Nanotechnology {\bf 11}, 387 (2000).

\bibitem{Hayashi} T. Hayashi, T. Fujisawa, H. D. Cheong, Y. H. Jeong,
and Y. Hirayama, Phys. Rev. Lett. {\bf 91}, 226804 (2003).

\bibitem{DiCarlo} L. DiCarlo, H. J. Lynch, A. C. Johnson, L. I. Childress,
K. Crockett, C. M. Marcus, M. P. Hanson, and A. C. Gossard,
Phys. Rev. Lett. {\bf 92}, 226801 (2004).

\bibitem{Petta} J. R. Petta, A. C. Johnson, C. M. Marcus, M. P. Hanson,
and A. C. Gossard, e-print cond-mat/0408139.

\bibitem{Schofield} S. R. Schofield, N. J. Curson, M. Y. Simmons,
F. J. Rue$\beta$, T. Hallam, L. Oberbeck, and R. G. Clark,
Phys. Rev. Lett. {\bf 91}, 136104 (2003).

\bibitem{Dzurak} A. S. Dzurak, L. C. L. Hollenberg, D. N. Jamieson,
F. E. Stanley, C. Yang, T. M. Buhler, V. Chan, D. J. Reilly, C. Wellard,
A. R. Hamilton, C. I. Pakes, A. G. Ferguson, E. Gauja, S. Prawer,
G. J. Milburn, and R. G. Clark, e-print cond-mat/0306265.

\bibitem{Hollenberg} L. C. L. Hollenberg, A. S. Dzurak, C. Wellard,
A. R. Hamilton, D. J. Reilly, G. J. Milburn, and R. G. Clark,
Phys. Rev. B {\bf 69}, 113301 (2004).

\bibitem{Kettle} L. M. Kettle, H.-S. Goan, S. C. Smith, C. J. Wellard,
L. C. L. Hollenberg, and C. I. Pakes, Phys. Rev. B {\bf 68}, 075317 (2003).

\bibitem{Barrett} S. D. Barrett and G. J. Milburn,
Phys. Rev. B {\bf 68}, 155307 (2003).

\bibitem{Fedichkin} L. Fedichkin and A. Fedorov,
Phys. Rev. A {\bf 69}, 032311 (2004).

\bibitem{Herring} C. Herring, Rev. Mod. Phys. {\bf 34}, 631 (1962).

\bibitem{NOTE}The resonant approximation is valid if the absolute values
of $\delta_L=\omega_L-(E_3-E_L)/\hbar$ and
$\delta_R=\omega_R-(E_3-E_R)/\hbar$ are small compared to
$(E_4-E_3)/\hbar$. One should keep in mind that the degeneracy of the Si
conduction band can result in the degenerate molecular state
$|\chi_3\rangle$ and the corresponding leakage of the quantum information
due to population of more than one excited states. Note, however, that this
degeneracy is lifted by the surface effects, the gate potentials, and the
internal strain due to donors (by analogy with the external stress effects
[B. Koiller, X. Hu, and S. Das Sarma, Phys. Rev. B {\bf 66}, 115201 (2002)]).
To find the spacing between the energy levels in the multiplet and thus to
quantify the allowed degree of deviation from the resonant conditions, one
should make numerical calculations for a specific donor configuration.
There is no such a problem in semiconductors with non-degenerate
conduction band edges.

\bibitem{Openov} L. A. Openov, Phys. Rev. B {\bf 60}, 8798 (1999);
A. V. Tsukanov and L. A. Openov,
Fiz. Tekh. Poluprovodn. (St. Petersburg) {\bf 38}, 94 (2004)
[Semiconductors {\bf 38}, 91 (2004)].

\bibitem{Brandes} T. Brandes and T. Vorrath,
Phys. Rev. B {\bf 66}, 075341 (2002).

\bibitem{Fermi} One can easily check that the Fermi golden rule is justified
for evaluation of the relaxation rate in the case under consideration, see
L. A. Openov, Phys. Rev. Lett. {\bf 93}, 158901 (2004).

\bibitem{DiVincenzo} D. P. DiVincenzo, Fortschr. Phys. {\bf 48}, 771 (2000).

\end{thebibliography}
\end{document}